\documentclass[amsmath,amssymb,aps,prx,reprint,
superscriptaddress
]{revtex4-2}
\usepackage{amsmath}
\usepackage[table]{xcolor}
\usepackage{amsfonts,amssymb}
\usepackage{graphicx}
\usepackage{hyperref}
\usepackage{academicons}
\usepackage{multirow}
\usepackage[version=4]{mhchem}
\usepackage{makecell}
\usepackage{pgf}
\usepackage{booktabs}
\usepackage[english]{babel}
\usepackage[normalem]{ulem}
\usepackage{natbib}
\usepackage{bm}

\newcommand{\GWunit}{GW\,m$^{-2}$\,K$^{-1}$}

\begin{document}

\title{Scalable long-range machine learning interatomic potentials with electrical response for atomistic material modeling
}

\author{Yicheng Chen}

\affiliation{Department of Materials Science and Engineering, National University of Singapore, Singapore}

\author{Xiaoyun Yu}
\thanks{These authors contributed equally.}
\affiliation{Department of Materials Science and Engineering, National University of Singapore, Singapore}

\author{Hyejeong Song}
\thanks{These authors contributed equally.}
\affiliation{Department of Materials Science and Engineering, National University of Singapore, Singapore}

\author{Xintian Wang}
\thanks{These authors contributed equally.}
\affiliation{Department of Materials Science and Engineering, National University of Singapore, Singapore}

\author{Zihui Li}
\thanks{These authors contributed equally.}
\affiliation{Department of Materials Science and Engineering, National University of Singapore, Singapore}

\author{Zixi Xia}
\affiliation{Department of Materials Science and Engineering, National University of Singapore, Singapore}
\affiliation{Department of Physics, National University of Singapore, Singapore}

\author{Junmin Chen}
\affiliation{Department of Materials Science and Engineering, National University of Singapore, Singapore}

\author{Peichen Zhong}
\email[]{zhongpc@nus.edu.sg}
\affiliation{Department of Materials Science and Engineering, National University of Singapore, Singapore}

\date{\today}

\begin{abstract}
Accurate atomistic simulations of materials under electric fields require long-range electrostatics and electrical-response modeling at length scales beyond the reach of conventional first-principles methods. Yet these capabilities remain difficult to reconcile with the scalability of machine-learning interatomic potentials (MLIPs). Strictly local models enable efficient spatial parallelization but neglect long-range interactions, whereas existing electrostatic extensions introduce global dependencies that hinder distributed computation. Here, we overcome this limitation by analytically separating the long-range electrostatic forces and Born effective charges (BECs) into a global electrostatics calculation and strictly local derivatives of learned latent charges. This formulation preserves the spatial decomposability of equivariant MLIPs while enabling distributed evaluation of energies, forces, stresses, and BECs across different boundary conditions. Implemented as DPA4C-LR on the \texttt{nvalchemi} platform, the framework scales to 5.2 million atoms on four GPUs at approximately twice the computational cost of its short-range backbone. Trained on energies and forces, DPA4C-LR accurately infers BECs and reproduces the field-dependent infrared spectra of water. We further demonstrate its capabilities by recovering the characteristic slim polarization hysteresis of relaxor \ce{Pb(Mg_{1/3}Nb_{2/3})O3}, which is obscured by finite-size effects in smaller simulations; predicting a room-temperature field-driven ionic conductivity of $6.8 \pm 0.7$~mS/cm in amorphous \ce{Li_{0.9}NbO_{0.9}Cl_{4.1}}; and modeling interfacial thermal transport across Si/diamond interfaces with slab-corrected electrostatics. By combining long-range physics, electrical-response inference, and distributed computation, this framework extends predictive atomistic modeling toward device-relevant length scales.
\end{abstract}


\maketitle

\section{Introduction}

Atomistic materials modeling increasingly needs to reach device scales while coupling thermal and electrical fields at \textit{ab initio} accuracy.
In an operating device, electric fields drive polarization switching and ion transport, and the heat dissipated by these processes must leave across internal interfaces \cite{maier_nanoionics_2005,giri_thermal_trans_2020,dawber_physics_2005}.
The thermal and electrical responses that emerge from atomic motion coupled to internal or applied electric fields lie far beyond the reach of brute-force electronic-structure calculations such as density functional theory (DFT) \cite{hohenberg_hk_1964,kohn_ks_1965}.
Predictive simulation of thermally and electrically coupled processes therefore imposes three requirements on an atomistic model: \textit{ab initio} accuracy, long-range electrostatics with an electric-field response, and device-relevant length and time scales.

Machine learning interatomic potentials (MLIPs) have emerged as the principal tool for bridging these spatial and temporal scales \cite{behler_bpnn_2007,bartok_gap_2010,zhang_deep_2018,schutt_schnet_2018,batzner_nequip_2022,batatia_mace_2022,passaro_escn_2023}.
For instance, foundation potentials (FPs) trained on extensive \textit{ab initio} datasets \cite{deng_chgnet_2023, barroso-luque_open_2024, levine_open_2025, kaplan_matpes_2025} enable zero-shot atomistic simulations, from structure relaxation to molecular dynamics (MD) \cite{chen_universal_2022, batatia_foundation_2025, wood_uma_2025, yuan_foundation_2026}.
However, the cost of device-level simulation grows with system size, and brute-force scaling of FPs may not be economically viable.
A practical alternative is a compact, specialized model distilled from an FP, optionally fine-tuned on DFT data, for the chemistry of interest; such a student model trades generality for speed at a modest loss of accuracy \cite{amin_towards_2025, wang_pre-training_2025}.

Strictly local MLIPs offer a way to speed up such models \cite{zhang_deep_2018, musaelian_learning_2023}. When the energy of an atom depends only on its environment within a finite cutoff, the simulation cell can be divided among many graphics processing units (GPUs), and each GPU then needs only a thin shell of ghost atoms from its neighbors.
Strictly local equivariant architectures exploit this property to reach throughputs close to those of empirical potentials. For example, DPA4C simulates more than two billion atoms on 1,024 GPUs \cite{li_dpa4c_2026}. With accuracy and scale within reach, the open question is whether long-range electrostatics and electrical response can be delivered at the same scale.

Locality has a price. A model with a finite cutoff cannot represent electrostatic interactions that decay as $1/r$ \cite{king_machine_2025}, which govern charged defects \cite{broberg_high_2023,ko_fast_2025}, polar interfaces \cite{batatia_macepolar_2026}, and the splitting between longitudinal and transverse optical phonons \cite{fan_qnep_2026}.
Several studies have shown that including long-range electrostatics in MLIPs improves fitting accuracy over short-range MLIPs while recovering the correct $1/r$ asymptotics \cite{ko_fourth-generation_2021, zhang_deep_2022}. 
A short-range model also offers no variable through which an external electric field can act.

The central quantity for electrical response is the Born effective charge (BEC) tensor,
\begin{equation}
    \mathbf{Z}^*_i = \frac{\partial \mathbf{P}}{\partial \mathbf{r}_i} = \left( \frac{\partial \mathbf{F}_i}{\partial \bm{\mathcal{E}}} \right)^\top
\end{equation}
which relates the response of the cell dipole moment $\mathbf{P}$ to atomic displacements $\mathbf{r}_i$ and, equivalently, the response of atomic forces $\mathbf{F}_i$ to an applied electric field $\bm{\mathcal{E}}$. 
The BEC provides access to infrared (IR) spectra and dielectric response, and enables MD under a finite field, from which ionic conductivity, polarization--electric-field ($P$--$\mathcal{E}$) hysteresis loops, and ferroelectric switching follow \cite{zhong_les_ir_2025,yu_bec_2025}.
Among the long-range approaches, latent Ewald summation (LES) is particularly economical \cite{cheng_les_2025}.
Atomic charges are latent variables learned from energies and forces alone, and one can infer polarization and BECs from these charges without additional labels.

Yet LES computes the long-range energy and the cell polarization by summing over every atom in the simulation cell, and obtains BECs by automatic differentiation of the polarization.
The computational graph of the whole cell must therefore reside in a single differentiation context, and the long-range augmentation does not inherit the domain decomposition of the backbone to which it is attached.
The charge-aware neuroevolution potential (qNEP) lifts both restrictions, the whole-cell sum and the whole-cell backward pass: with latent charges that depend only on the local environment, analytically derived forces, and a real-space BEC expression, it reaches million-atom simulations with electrical response on a single GPU \cite{fan_qnep_2026}.
However, qNEP accuracy is limited by its descriptor, yielding markedly higher errors than equivariant models \cite{ying_nep_2025,liang_nep89_2026,li_dpa4c_2026,fan_qnep_2026}. Conversely, the fastest strictly local equivariant models contain no electrostatics and no response to external fields.

In this work, we scale long-range MLIPs with electrical-response inference to multi-million-atom systems across GPUs while retaining the accuracy of an equivariant backbone.
When latent charges are read out from a local descriptor, the only global operation is a classical electrostatics solve over charges and positions, for which parallel $O(N \log N)$ mesh algorithms are mature.
We therefore apply the chain rule for the long-range energy analytically instead of tracing it through the whole cell, so that energies, forces, stresses, and BECs decompose over spatial partitions in the same way as in a short-range model.
This meets the three requirements above: equivariant accuracy from the backbone, long-range electrostatics and field response from the latent charges, and scale from a domain decomposition that the long-range term now shares.
We implement this approach for non-periodic systems, three-dimensional periodic boundary conditions (PBC), and interface slabs with dipole correction.
The resulting model, termed DPA4C-LR, combines the DPA4C backbone with particle mesh Ewald (PME) electrostatics \cite{darden_pme_1993,essmann_spme_1995,loche_pme_2025} on the NVIDIA ALCHEMI Toolkit-Ops (\texttt{nvalchemi}) platform and runs across multiple GPUs \cite{nvidia_alchemi_toolkit_ops}.
On this platform, the model reaches 5.2 million atoms on four GPUs, with 43\% parallel efficiency measured at the largest box all GPU counts share (1.96 million atoms) and a long-range overhead of 2.1$\times$ over the short-range backbone (based on four GPUs).
We then demonstrate three applications that each require large cells: chemical order/disorder in a relaxor ferroelectric, ionic transport in an amorphous electrolyte, and heat transport across a solid--solid interface.
The relaxor ferroelectric \ce{Pb(Mg_{1/3}Nb_{2/3})O3} requires coupling to an applied field in cells large enough to contain polar nanoregions to obtain the $P$--$\mathcal{E}$ hysteresis loops.
Ionic transport in off-stoichiometric \ce{Li_{0.9}NbO_{0.9}Cl_{4.1}} (LNOC) solid-state electrolytes requires sufficient charge-carrier diffusion events in an amorphous structure.
Thermal transport across silicon/diamond interfaces requires slab boundary conditions and long simulation cells.

\begin{figure*}[tbp]
    \centering
    \includegraphics[width=\linewidth]{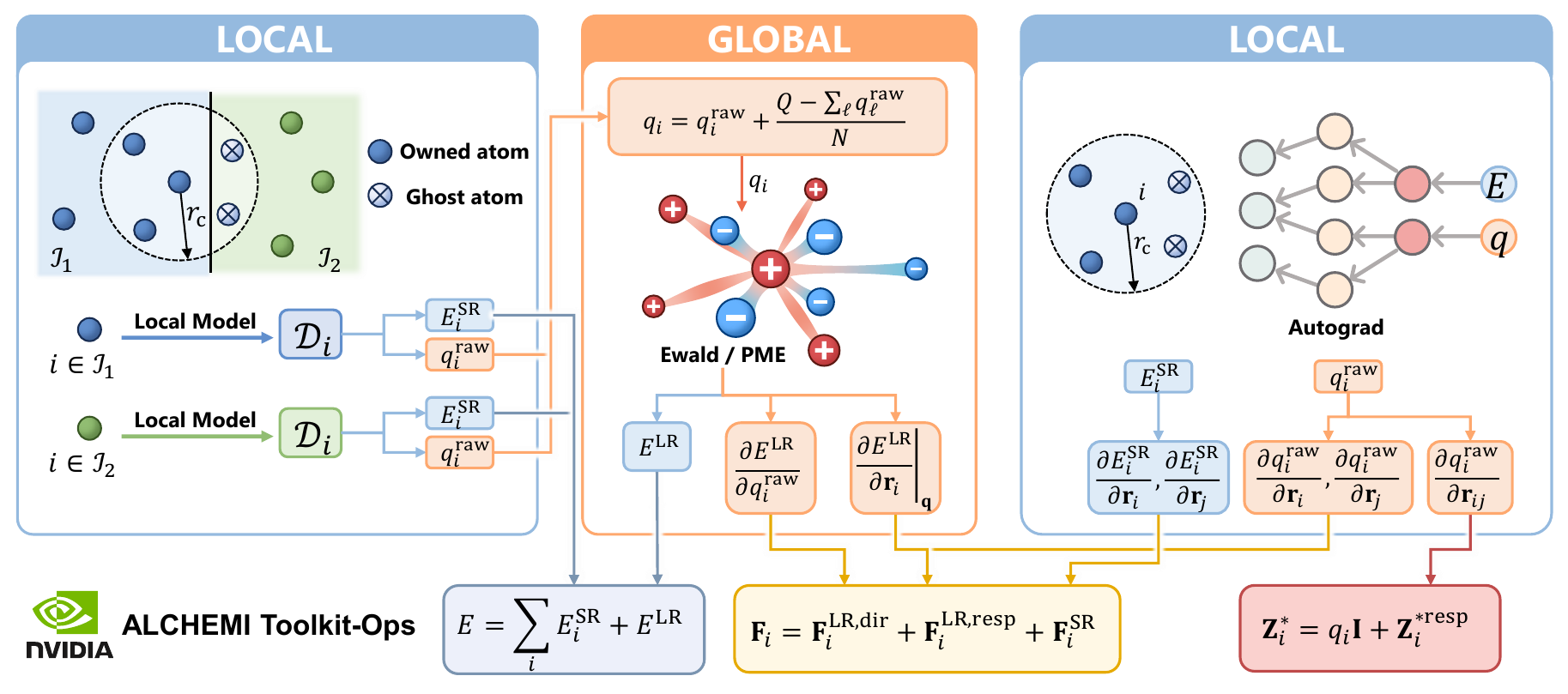}
    \caption{
        Schematic of the distributed workflow for evaluating energies, forces, and Born effective charges (BECs) with strictly local models augmented by long-range electrostatics. Local models predict per-atom short-range energies and raw charges on each partition. A global Ewald/PME step, implemented with the NVIDIA ALCHEMI Toolkit-Ops, then enforces the total-charge constraint and evaluates the long-range energy and its derivatives, while local automatic differentiation provides the remaining position derivatives. These contributions are assembled into the total energy, forces, and BECs.
    }
    \label{fig:workflow}
\end{figure*}

\section{Methods}\label{sec:methods}

\subsection{Model design}

We employ DPA4C \cite{li_dpa4c_2026} as the local representation backbone. DPA4C is an equivariant graph neural network with a single message-passing layer, yielding strictly local atomic representations that can be evaluated independently across spatial partitions. For each atom $i$, the local environment is mapped to a descriptor $\mathbf{D}_i$,
\begin{equation}
    \mathbf{D}_i = \mathcal{F} \left( Z_i, \mathbf{r}_i, \left\{ Z_j, \mathbf{r}_j \right\}_{j \in \mathcal{N}_i} \right)
\end{equation}
where $\mathcal{N}_i$ denotes the set of neighboring atoms within the cutoff radius, $Z_i$ is the atomic number, and $\mathbf{r}_i$ the atomic position.

Following the LES framework \cite{cheng_les_2025,kim_les_2025}, the short-range energy and the raw latent charge are predicted by separate fitting networks:
\begin{equation}
    E^\text{SR} = \sum_i \phi_e \left( \mathbf{D}_i \right),\quad q^\text{raw}_i = \phi_q \left( \mathbf{D}_i \right)
\end{equation}
To ensure consistency with the prescribed total charge $Q$ of the system, the raw charges are subsequently projected as
\begin{equation}
    q_i = q^\text{raw}_i
        + \frac{Q - \sum_{\ell=1}^{N} q^\text{raw}_\ell}{N}
\end{equation}
where $N$ is the number of atoms. This projection guarantees the exact global charge conservation, $\sum_i q_i = Q$.

The Coulomb interaction is separated into short- and long-range contributions using Gaussian screening:
\begin{equation}
    \frac{1}{r} =\frac{\text{erfc}\left( \alpha r \right)}{r}
                     + \frac{\text{erf}\left( \alpha r \right)}{r}
                  \label{eq:coul_split}
\end{equation}
where $\alpha$ denotes the Gaussian screening parameter, while $\operatorname{erf}(\cdot)$ and $\operatorname{erfc}(\cdot)$ denote the error function and its complementary counterpart, respectively. 
The rapidly decaying short-range contribution is absorbed into $E^\text{SR}$, while the smooth long-range contribution is evaluated explicitly from the latent charges.

Depending on the boundary conditions, the long-range energy is evaluated directly in real space for nonperiodic systems, through reciprocal-space Ewald summation or PME for 3D periodic systems, and with an additional Yeh–Berkowitz (YB) correction for 2D periodic slabs \cite{essmann_spme_1995,yeh_berkowitz_1999}. The numerical formulations are detailed in the Supplementary Information (SI).

\subsection{Distributed parallelization}

The parallel implementation exploits the strict locality of DPA4C to spatially decompose the short-range calculations, as illustrated in Fig.~\ref{fig:workflow}. The simulation cell is divided into $P$ spatial partitions, each assigned to one device. Partition $p$ contains owned atoms $\mathcal{I}_p$ and ghost atoms $\mathcal{G}_p$, which provide the complete local environments within the model cutoff.

The short-range atomic energies and raw latent charges are evaluated independently on each partition. The projected charges are then used for global long-range electrostatic evaluation, yielding the total energy
\begin{equation}
    E = \sum_{p=1}^P \sum_{i \in \mathcal{I}_p} E^\text{SR}_i + E^\text{LR}
\end{equation}

For the short-range forces, atomic energies are accumulated over owned atoms, while their derivatives include contributions from both owned and ghost atoms. The resulting force contributions are summed across partitions according to the global atom indices:
\begin{equation}
\begin{aligned}
    \mathbf{F}^\text{SR}_{j,p} &= -\sum_{i \in \mathcal{I}_p}
                                 \frac{\partial E^\text{SR}_i}{\partial \mathbf{r}_j},
                                 \quad j \in \mathcal{I}_p \cup \mathcal{G}_p \\
    \mathbf{F}^\text{SR}_j &= \sum_{p=1}^P \mathbf{F}^\text{SR}_{j,p}
\end{aligned}
\end{equation}

The long-range forces contain both direct electrostatic and charge-response contributions:
\begin{equation}
\begin{aligned}
    \mathbf{F}^\text{LR}_j &= - \frac{d E^\text{LR}}{d \mathbf{r}_j} \\
    &= - \left. \frac{\partial E^\text{LR}}{\partial \mathbf{r}_j} \right|_\mathbf{q}
       - \sum_i \frac{\partial E^\text{LR}}{\partial q^\text{raw}_i} \frac{\partial q^\text{raw}_i}{\partial \mathbf{r}_j} \\
    &= \mathbf{F}^\text{LR,dir}_j + \mathbf{F}^\text{LR,resp}_j
\end{aligned}
\end{equation}

The direct electrostatic forces and charge gradients are obtained from the global long-range calculation, whereas the coordinate derivatives of the latent charges are evaluated locally through automatic differentiation. The resulting short-range and charge-response force contributions are accumulated across partitions using the corresponding global atom indices. Stress tensors are evaluated analogously (see Supplementary Information).

\subsection{Born effective charges}
Following \citet{zhong_les_ir_2025}, the latent charges are rescaled to account for electronic screening before evaluating the macroscopic polarization and BECs:
\begin{equation}
    \tilde{q}_i = \sqrt{\varepsilon_\infty} q_i
\end{equation}
where $\varepsilon_\infty$ denotes the high-frequency relative permittivity. We follow the BEC expression of \citet{fan_qnep_2026}, which decomposes $\mathbf{Z}^*_i$ into a direct charge contribution and a charge-response contribution:
\begin{equation}
    \begin{aligned}
    \mathbf{Z}^*_i &= \tilde{q}_i \mathbf{I} + \frac{1}{2}\sum_{j\neq i}\left(
                      \mathbf{r}_{ij} \otimes \frac{\partial \tilde{q}_i}{\partial \mathbf{r}_{ij}} 
                    - \mathbf{r}_{ji} \otimes \frac{\partial \tilde{q}_j}{\partial \mathbf{r}_{ji}}
                 \right) \\
                 &= \mathbf{Z}^{*,\text{dir}}_i + \mathbf{Z}^{*,\text{resp}}_i
    \end{aligned} \label{eq:BEC}
\end{equation}
where $\mathbf{I}$ is the $3\times3$ identity matrix, $\mathbf{r}_{ij} = \mathbf{r}_j - \mathbf{r}_i$ is the vector from atom $i$ to atom $j$, and $\otimes$ denotes the outer product.

For a configuration-independent screening factor, the global charge-projection term cancels from the pairwise response expression, yielding
\begin{equation}
    \mathbf{Z}^{*,\text{resp}}_i = \frac{\sqrt{\varepsilon_\infty}}{2}\sum_{j\neq i}\left(
                      \mathbf{r}_{ij} \otimes \frac{\partial q^\text{raw}_i}{\partial \mathbf{r}_{ij}} 
                    - \mathbf{r}_{ji} \otimes \frac{\partial q^\text{raw}_j}{\partial \mathbf{r}_{ji}}
                 \right)
\end{equation}
The raw-charge derivatives are strictly local and can therefore be evaluated independently within each spatial partition, followed by a scatter-sum over global atom indices. The detailed derivation is provided in the SI.

\section{Results}

\subsection{Bulk water}
\begin{figure*}[htbp]
    \centering
    \includegraphics[width=0.8\linewidth]{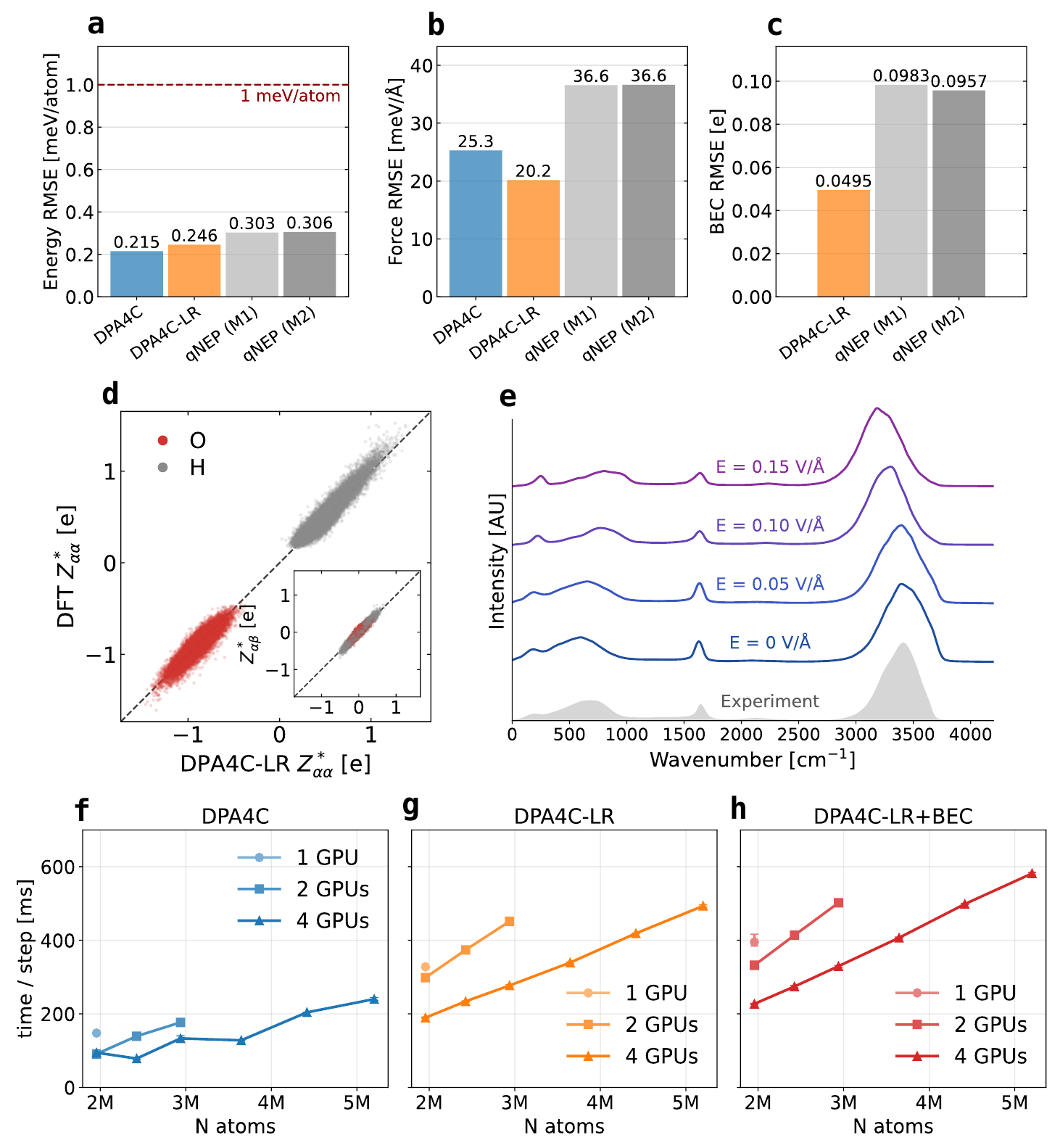}
    \caption{
        DPA4C-LR for bulk water.
        (a)--(c)~Test-set RMSEs of energies, forces, and BECs for DPA4C, DPA4C-LR, and the M1 and M2 variants of qNEP. 
        (d)~Diagonal BECs ($Z^*_{\alpha\alpha}$) predicted by DPA4C-LR against DFPT reference values, with the off-diagonal components ($Z^*_{\alpha\beta}$) in the inset.
        (e)~Simulated IR spectra of bulk water under static electric fields of 0--0.15~V/{\AA}, compared with experiment \cite{bertie_waterir_1996}.
        (f)--(h)~Time per MD step against number of atoms on one, two, and four NVIDIA H200 GPUs for DPA4C, DPA4C-LR, and DPA4C-LR with on-the-fly BEC evaluation.
    }
    \label{fig:water}
\end{figure*}

We first examine bulk water as a representative condensed-phase system in which long-range electrostatics and their response to atomic displacements are essential. Following the benchmark protocol of Ref.~\cite{zhong_les_ir_2025}, we trained DPA4C-LR on the same bulk-water dataset, using only energies and forces as training targets and without explicit supervision of atomic charges or BECs.
For comparison, we also trained the original DPA4C model and the M1 and M2 variants of qNEP \cite{fan_qnep_2026} on the same dataset.

Figure~\ref{fig:water}a--c shows the root-mean-square errors (RMSEs) of the predicted energies, forces, and BECs on the test set. DPA4C-LR achieves an energy RMSE comparable to that of the baseline DPA4C model (0.25 vs. 0.22~meV/atom) while reducing the force RMSE from 25.3 to 20.2~meV/{\AA}. The force error is also well below that of the M1 and M2 variants of qNEP (both 36.6~meV/{\AA}).
For the BEC evaluation, following the protocol of Ref.~\cite{fan_qnep_2026}, we computed reference BECs by density functional perturbation theory (DFPT) \cite{baroni_dfpt_2001} for 65 configurations (10\% of the training set) and obtained the high-frequency relative permittivity $\varepsilon_\infty=3.07$ as the least-squares slope of the DFPT BECs against the DPA4C-LR charges. 
As shown in Fig.~\ref{fig:water}d, the resulting BECs are more accurate than those of qNEP, with an RMSE of 0.05 versus 0.10~$e$.

The IR spectra were computed from MD simulations. We performed NVT MD at 300~K with a Nos\'e--Hoover thermostat \cite{nose_molecular_1984,hoover_canonical_1985,martyna_nhc_1992}, starting from the same initial configuration as in Ref.~\cite{zhong_les_ir_2025}. With a time step of 0.25~fs, each trajectory was propagated for 1~ns under a static electric field of 0, 0.05, 0.10, or 0.15~V/{\AA} applied along the $x$ direction. The polarization current, $\mathbf{J} (t) = \sum_i \mathbf{Z}^*_i (t) \cdot \mathbf{v}_i (t)$, where $\mathbf{v}_i$ is the atomic velocity, was recorded every 0.5~fs. The IR spectrum was then obtained from the Fourier transform of the polarization-current autocorrelation function,
\begin{equation}
    I\left(\omega\right) \propto \int_0^T dt \left< \mathbf{J}(0) \cdot \mathbf{J}(t) \right> \text{e}^{-\text{i}\omega t }.
\end{equation}

Figure~\ref{fig:water}e shows the IR spectra simulated with DPA4C-LR. The positions and shapes of the intramolecular bands agree with experiment \cite{bertie_waterir_1996}, and the low-frequency librational band and the hydrogen-bond translational band are also reproduced. Under static external fields, the band shifts match those reported in Ref.~\cite{zhong_les_ir_2025}. DPA4C-LR thus reproduces the IR spectrum of bulk water and its field dependence from a model trained solely on energies and forces.

To demonstrate the scalability of our framework, we benchmarked DPA4C, DPA4C-LR, and DPA4C-LR with BEC evaluation on up to four NVIDIA H200 GPUs. All models were evaluated in their compressed form using spatial decomposition with one domain per GPU. The scaling performance is summarized in Fig.~\ref{fig:water}f–h. 
For a 1.96-million-atom water system, scaling from one to four GPUs reduced the time per MD step from 148 to 95~ms for DPA4C, from 328 to 190 ms for DPA4C-LR (without BEC inference), and from 395 to 227 ms for DPA4C-LR+BEC (with on-the-fly BEC inference), corresponding to speedups of 1.56$\times$, 1.73$\times$, and 1.74$\times$, respectively. The long-range models maintained nearly constant four-GPU throughput as the system size increased, reaching 5.2 million atoms at 493 ms/step for DPA4C-LR and 582 ms/step with BEC evaluation. These results demonstrate that both long-range electrostatics and on-the-fly BEC evaluation remain scalable and computationally tractable at the multi-million-atom scale.

\begin{figure*}[htbp]
    \centering
    \includegraphics[width=0.9\linewidth]{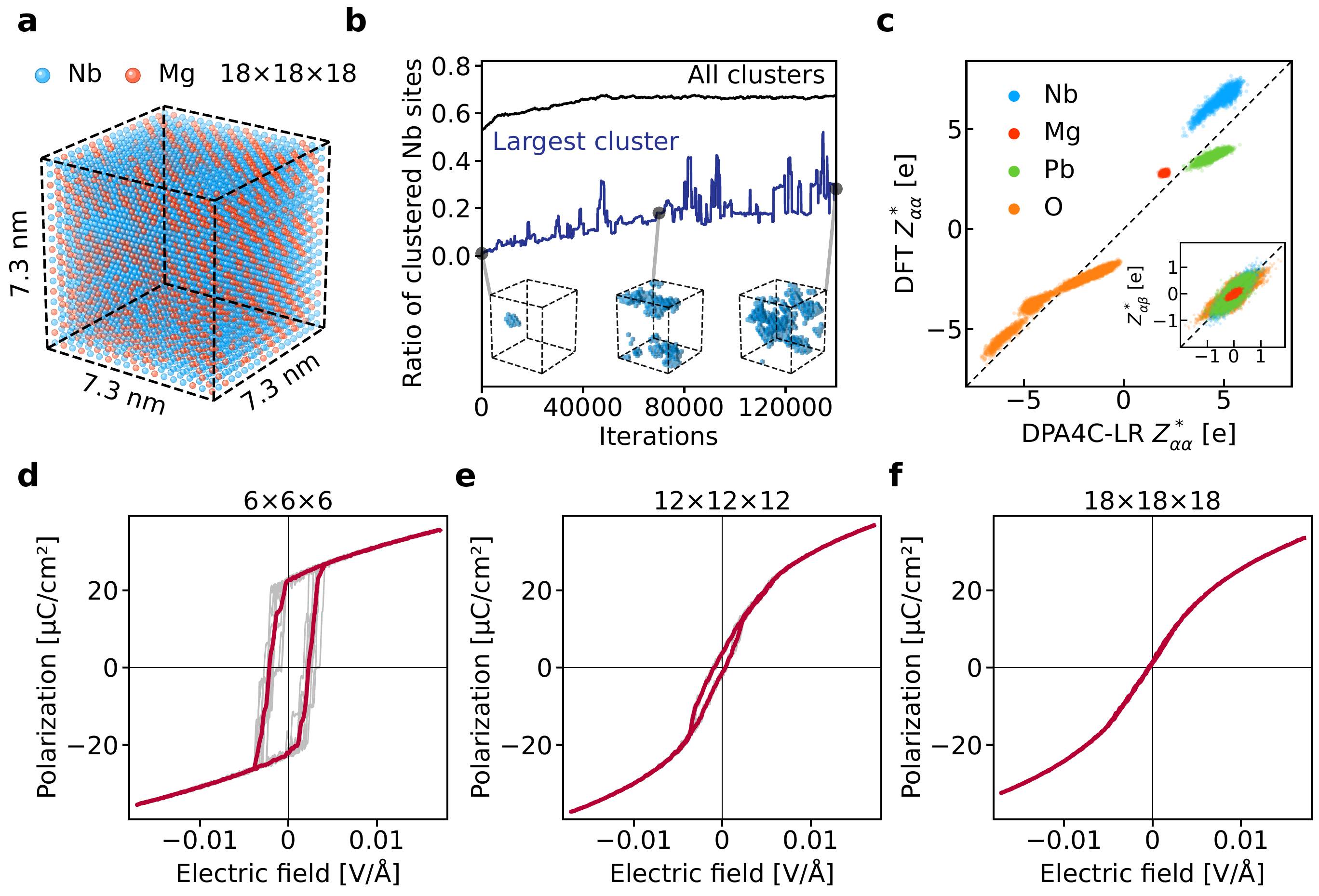}
    \caption{DPA4C-LR for relaxor PMN. (a)~$18\times18\times18$ PMN supercell from FIRE-Swap, showing only the B sites, with Nb in blue and Mg in red. (b)~Ratio of Nb sites in all Nb clusters (black) and in the largest cluster (navy) against the number of MC swap iterations. The insets show the largest cluster after 0, $7\times10^4$, and $1.4\times10^5$ iterations. (c)~Diagonal BECs ($Z^*_{\alpha\alpha}$) predicted by DPA4C-LR against DFPT, with the off-diagonal components ($Z^*_{\alpha\beta}$) in the inset. (d)--(f)~$P$--$\mathcal{E}$ loops along [111] at 300~K for $6\times6\times6$, $12\times12\times12$, and $18\times18\times18$ supercells. Red lines give the mean over replicas, and gray lines give the individual replicas.
    }
    \label{fig:pmn}
\end{figure*}

\subsection{\ce{Pb(Mg_{1/3}Nb_{2/3})O3} relaxor ferroelectrics}
With BEC inferences, one can simulate the $P$--$\mathcal{E}$ relation for ferroelectric materials. Unlike the ordered \ce{PbTiO3} case shown in Ref.~\cite{zhong_les_ir_2025}, relaxor ferroelectrics \ce{Pb(Mg_{1/3}Nb_{2/3})O3} (PMN) exhibit complex intrinsic chemical order in which Mg and Nb form a mesh-grid of locally ordered domains, while the Nb-rich domains interconnect to form a percolating network \cite{xu_intrinsic_2026}. This structural complexity extends over nanometer length scales and is experimentally characterized by B-site cation ordering and local oxygen-octahedral distortions, which disrupt long-range polarization order and give rise to relaxor behavior~\cite{kumar_atomic_2021}. This complex partially ordered structure requires a large-scale supercell representation to capture chemical ordering and to compute the $P$--$\mathcal{E}$ hysteresis loop without finite-size effects.

We trained DPA4C-LR for PMN by distillation from the fine-tuned DPA-3 model of Ref.~\cite{xu_intrinsic_2026} and evaluated it against 200 PBEsol \cite{perdew_pbesol_2008} DFT configurations (see SI).
Against DFT, DPA4C-LR reaches energy and force RMSEs of 0.36~meV/atom and 40.3~meV/{\AA}, compared with 0.43~meV/atom and 46.9~meV/{\AA} for DPA4C.
To obtain the B-site ordering, we ran swap Monte Carlo (MC) with the FIRE-Swap method at an effective temperature $T_{\text{eff}}=600$~K \cite{xu_intrinsic_2026}.
Figure~\ref{fig:pmn}a shows a representative $18\times18\times18$ supercell of the five-atom perovskite ABO$_3$ unit cell (29,160 atoms, 5,832 B sites).
Consistent with Ref.~\cite{xu_intrinsic_2026}, the supercell accommodates chemically ordered Nb--Mg domains and Nb-rich domains.
Figure~\ref{fig:pmn}b shows the formation of Nb clusters during the MC simulation in the $18\times18\times18$ supercell. The ratio of clustered Nb sites, normalized by one third of the number of B sites, rises from 0.53 to 0.67 within $4\times10^4$ iterations.
The largest Nb cluster grows from a ratio of 0.01 to 0.18 after $7\times10^4$ iterations and to 0.28 after $1.4\times10^5$ iterations.
The insets in Fig.~\ref{fig:pmn}b show the growth of the largest Nb cluster.

Figure~\ref{fig:pmn}c compares the BECs inferred by DPA4C-LR with those from DFPT for 30 PMN configurations of 135 atoms. The correlation coefficient over all diagonal components is 0.97. As PMN is a complex oxide, the BEC inference agrees qualitatively with the DFPT reference for the diagonal components. The off-diagonal components agree with DFPT within an RMSE of 0.22~$e$. A model trained only on energies and forces thus infers BECs that track the DFPT values from one local environment to the next.

Using the inferred BECs, we obtained $P$--$\mathcal{E}$ hysteresis loops from NVT MD at 300~K under an electric field along [111]. We compare $6\times6\times6$ (1,080 atoms), $12\times12\times12$ (8,640 atoms), and $18\times18\times18$ (29,160 atoms) supercells, each taken from a FIRE-Swap simulation of that size. Each cell was equilibrated at 300~K, first by NPT MD to set the lattice constant ($a=4.054$~\AA) and then by NVT MD for 1~ns with a Nos\'e--Hoover thermostat~\cite{nose_molecular_1984,hoover_canonical_1985,martyna_nhc_1992}. The field followed a triangular wave with a period of 1.6~ns and a physical amplitude of 0.017~V/\AA. The polarization was computed from the BEC-weighted displacements relative to the ideal cubic structure, $\boldsymbol{\mathcal{P}} = \frac{1}{V}\sum_i \mathbf{Z}^*_i \cdot \Delta\mathbf{r}_i$. 
Because the MD used unscaled BECs, we converted the field and polarization to physical units with $\sqrt{\varepsilon_\infty}=3.80$, obtained from the DFPT comparison of Fig.~\ref{fig:pmn}c in the same way as for water.

Figure~\ref{fig:pmn}d--f shows that the hysteresis shrinks as the supercell grows. The $6\times6\times6$ cell gives a square, ferroelectric-like loop with a remanent polarization $P_r$ of 22.4~$\mu$C/cm$^2$, and its polarization reverses in abrupt steps, as if the whole cell were a single domain.
The $12\times12\times12$ cell gives a narrower loop, with $P_r=2.5~\mu$C/cm$^2$ and a coercive field of about $\mathcal{E}_c=6.4\times10^{-4}$~V/\AA.
In the $18\times18\times18$ cell, the polarization follows the field continuously and the loop almost closes.
The polarization at the maximum field is similar for all three sizes, between 33 and 37~$\mu$C/cm$^2$.
The largest cell thus reproduces the slim, nearly anhysteretic loop that characterizes PMN as a relaxor ferroelectric at room temperature, consistent with the vanishing remanent polarization of PMN ceramics at 300~K~\cite{pandey_pmn_2018}. The loop shape is therefore a finite-size-sensitive observable: cells of tens of thousands of atoms are needed to recover the relaxor response, a scale at which DPA4C-LR runs routinely.

\subsection{Ionic transport in an amorphous solid electrolyte}

\begin{figure*}[htbp]
    \centering
    \includegraphics[width=\linewidth]{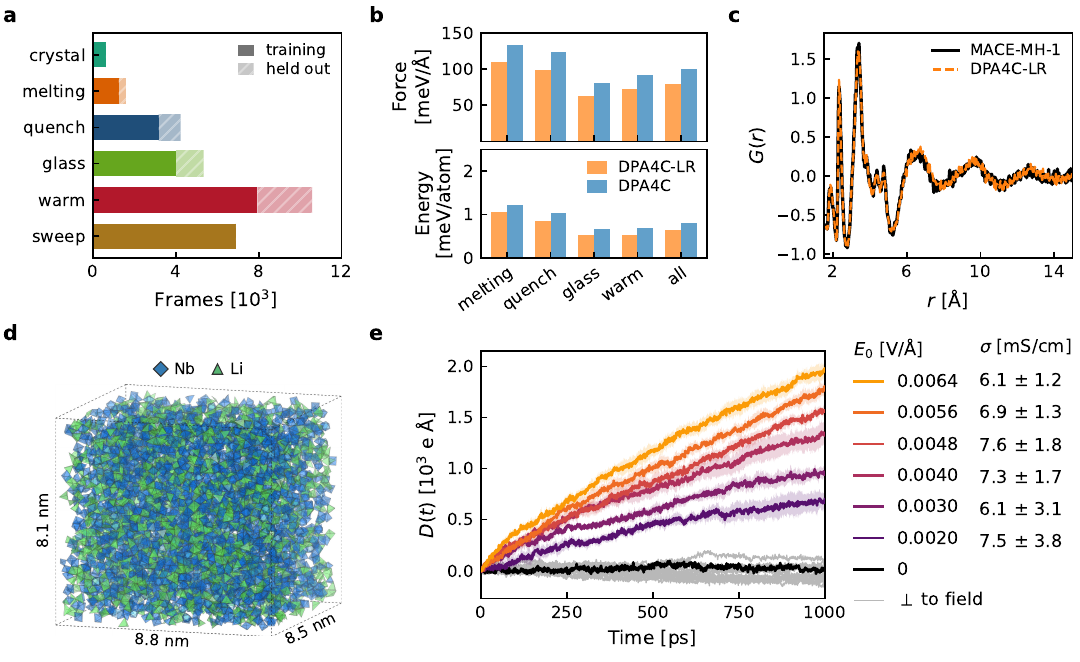}
    \caption{DPA4C-LR for amorphous LNOC.
    (a)~Training and held-out frames in each configuration family.
    (b)~Force and energy RMSEs of DPA4C-LR and the short-range DPA4C on the held-out frames, by family.
    (c)~Neutron-weighted $G(r)$ of a 2,236-atom glass quenched with MACE-MH-1 and with DPA4C-LR.
    (d)~One of the four 19,872-atom glasses quenched with DPA4C-LR.
    (e)~Charge displacement $D(t)$ at 300~K under static electric fields.
    Colored lines give the component along the field, averaged over the four glasses, with a band of one standard error between glasses.
    The black line gives the component along $x$ without a field (zero field), and gray lines give the components perpendicular to the field.
    The legend gives the physical field $\mathcal{E}$ (labeled $E_0$ in the plot) and the conductivity with its 95\% confidence interval over 100~ps blocks of the 500--1000~ps window.} 
    \label{fig:lnoc}
\end{figure*}

We next examine ionic transport in amorphous \ce{Li_{0.9}NbO_{0.9}Cl_{4.1}} (LNOC).
Mechanochemically prepared LNOC is about half amorphous \cite{newnhamCorrelationCoherenceLength2025}, and \citet{hussainExploringSuperionicConduction2024} proposed that the amorphous phase contributes to the conduction in lithium oxyhalides.
Because DPA4C-LR infers BECs on the fly, an electric field can be applied directly in MD, and we computed the Li conductivity of the glass at 300~K from MD under static fields in 19,872-atom cells.

The training data came from MD with the \texttt{matpes\_r2scan} head of MACE-MH-1 \cite{batatia_macemh_2025} in a 662-atom cell.
This cell is a $6\times2\times2$ supercell of the $Cc$ crystal \cite{singhCriticalRoleFramework2024} with ten Li vacancies and ten O atoms replaced by Cl, matching the experimental composition.
The trajectories cover the crystal, the heating ramp, the liquid at 1800~K, quenches at 1, 2, and 5~K/ps, the glass at 300~K, and the glass at 400--600~K.
A separate sweep adds liquids at 1200--3500~K and liquids compressed by 2 and 4~\% in cell edge, which bring atoms closer together and sample the short-bond environments needed for a robust potential.
The sweep also adds crystals at 450--1050~K (Table~S3 and Fig.~S1).
Figure~\ref{fig:lnoc}a shows the number of frames in each family.
We held out one melt and its seven glasses, 5,390 of the 29,257 labeled frames, as the test set.
Both DPA4C-LR and DPA4C were trained on the energies and forces of the remaining frames, without supervision of charges or BECs.
No DFT data were used in training; \citet{wang_practical_2026} showed that MACE-MH-1 melt-quench dynamics reproduce the experimental local structure of LNOC.

Figure~\ref{fig:lnoc}b compares the RMSEs of the DPA4C and DPA4C-LR models on the held-out frames.
DPA4C-LR exhibits lower energy and force errors in every family in Fig.~\ref{fig:lnoc}b.
Across all held-out frames, DPA4C-LR reaches an energy RMSE of 0.65~meV/atom and a force RMSE of 79~meV/{\AA} relative to the MACE-MH-1 labels; the short-range DPA4C reaches 0.81~meV/atom and 100~meV/{\AA} on the same frames.
We then quenched a 2,236-atom glass with DPA4C-LR and with MACE-MH-1 under the same melt-quench protocol. 
The smaller cell was chosen for computational efficiency with foundation potentials.
Figure~\ref{fig:lnoc}c compares the neutron-weighted pair distribution functions $G(r)$ of the two glasses (Eqs.~S35--S37).
The two curves differ by less than 0.2~{\AA}$^{-2}$ over 1.55--15~{\AA}, and the densities of the two glasses are 2.041 (DPA4C-LR) and 2.020~g/cm$^3$ (MACE-MH-1).

With DPA4C-LR, we quenched four 19,872-atom glasses from one 1800~K melt (Fig.~\ref{fig:lnoc}d and Fig.~S8a), with densities of 2.022--2.043~g/cm$^3$, only 0.1--1.1~\% above the MACE-MH-1 glass at 2.020~g/cm$^3$.
On each glass, we performed NVT MD at 300~K with a Nos\'e--Hoover thermostat \cite{nose_molecular_1984, hoover_canonical_1985, martyna_nhc_1992} and a time step of 1~fs.
After 10~ps of equilibration with the field on, each trajectory was propagated for 1~ns under a static field along $x$.
The MD used the unscaled BECs, so the physical field $\mathcal{E}$ is the applied field divided by $\sqrt{\varepsilon_\infty} = 2.49$ (Fig.~S2).
We ran MD at 6 physical fields from 0.0020 to 0.0064~V/{\AA} within the linear response, as shown in Fig.~\ref{fig:lnoc}e.
One MD step of a 19,872-atom glass takes 8.8~ms on one H200 GPU, corresponding to 10~ns/day (Fig.~S8b).

Figure~\ref{fig:lnoc}e plots the charge displacement $D_\alpha$, and the conductivity follows from its slope along the field,
\begin{equation}
D_\alpha(t) = \sum_i q_i\,\big[r_{i\alpha}(t) - r_{i\alpha}(0)\big],\quad \sigma = \frac{\dot{D}_x}{V\mathcal{E}},
\label{eq:lnoc_sigma}
\end{equation}
where $q_i$ is the nominal charge of atom $i$ ($+1$ for Li, $+5$ for Nb, $-2$ for O, and $-1$ for Cl) following Ref.~\cite{zhong_les_ir_2025}, $r_{i\alpha}$ is its position along direction $\alpha$, $\dot{D}_x = \Delta D_x/\Delta t$ is the slope of $D_x$ against time, $V$ is the cell volume, and $\mathcal{E}$ is the physical field.
At every field, $D_x$ grows with time and its slope increases with the field, whereas the perpendicular components stay near zero.
We divided the second half of each trajectory (500--1000~ps), over which the slope has settled, into five non-overlapping 100~ps blocks and fitted $\dot{D}_x$ by least squares in each block.

At each field, we report the mean over the blocks of all runs (20 blocks: four glasses with five blocks each) with its 95\% confidence interval $t\,s/\sqrt{n}$.
Here $s$ is the standard deviation of the $n$ block values, and $t$ is the Student's $t$ value for $n-1$ degrees of freedom (2.09 for $n = 20$).
The standard error does not increase when the block length is increased from 100 to 250~ps, so the blocks are effectively independent, and an analysis of variance detects no difference between the four glasses (see SI Sec.~IV~C).
Over 0.0020--0.0064~V/{\AA}, the six fields give ionic conductivities $\sigma$ between 6.1 and 7.6~mS/cm.
These values agree within their 95\% confidence intervals, consistent with linear response.
Their inverse-variance-weighted mean is $\sigma = 6.8 \pm 0.7$~mS/cm (95\% confidence interval, $t = 1.99$).

The zero-field runs provide an independent estimate of $\sigma$.
Without a field, the conductivity is instead estimated from the thermal fluctuations of $D_\alpha$ through the Einstein--Helfand~\cite{helfand_transport_1960} relation,
\begin{equation}
\sigma = \frac{1}{2Vk_\mathrm{B}T}\,\frac{\Big\langle \big[D_\alpha(t_0+\tau) - D_\alpha(t_0)\big]^2 \Big\rangle}{\tau},
\label{eq:lnoc_eh}
\end{equation}
where $k_\mathrm{B}T$ is the thermal energy at 300~K.
To match the window of the field-driven estimate, we evaluated Eq.~\eqref{eq:lnoc_eh} with $t_0 = 500$~ps and $\tau = 500$~ps.
Because no direction is preferred without a field, the three components $\alpha=x,y,z$ of each of the four glasses give 12 independent samples.
Their mean is $\sigma = 7 \pm 5$~mS/cm (95\% confidence interval, $t = 2.20$).
This interval contains the field-driven value of 6.8~mS/cm but is seven times wider ($\pm 5$ versus $\pm 0.7$~mS/cm), because the equilibrium fluctuations of $D_\alpha$ are small compared with the field-induced drift.
Measured conductivities of \ce{Li_{1-x}NbO_{1-x}Cl_{4+x}} at 25~$^\circ$C range from 1.7 to 9.4~mS/cm for $x$ from $-0.10$ to 0.20 \cite{newnhamCorrelationCoherenceLength2025}.
The field-driven value for the glass, $\sigma = 6.8 \pm 0.7$~mS/cm, lies inside this range.
The comparison is qualitative: the measured samples are about half amorphous, and their impedance spectra do not separate bulk from grain-boundary resistance \cite{newnhamCorrelationCoherenceLength2025}.
For this glass, field-driven MD with on-the-fly BEC inference thus gives $\sigma$ with a seven-fold narrower confidence interval than the Einstein--Helfand estimate from trajectories of the same length.

\begin{figure*}[htbp]
    \centering
    \includegraphics[width=\linewidth]{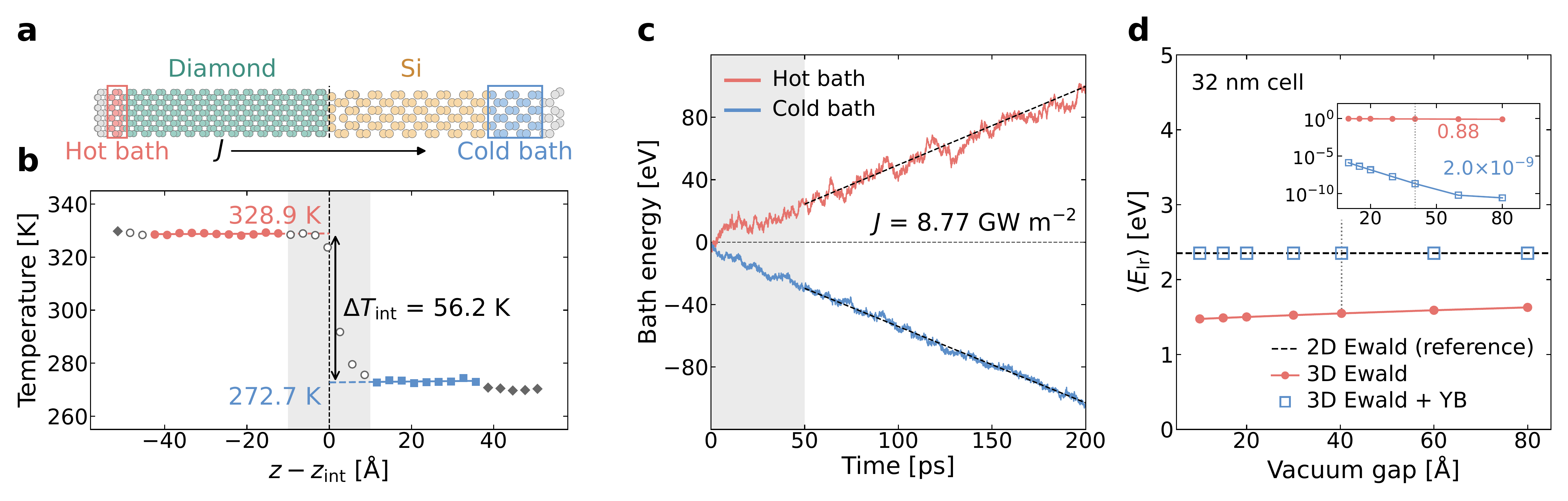}
    \caption{
    (a)~NEMD setup for the 11~nm Si/diamond interface.
    (b)~Temperature profile from one representative production run, with linear extrapolations to determine the interfacial temperature drop $\Delta T_{\mathrm{int}}$.
    (c)~Cumulative energy exchanged with the thermostats in the same run, from which the heat flux $J$ is obtained.
    (d)~Long-range electrostatic energy of the 32~nm cell as a function of vacuum gap, comparing direct 2D Ewald summation with uncorrected and Yeh--Berkowitz-corrected 3D Ewald summation. The dotted line marks the production vacuum gap of 40.3~{\AA}; the inset shows the maximum absolute deviation from the 2D reference.}
    \label{fig:nemd}
\end{figure*}

\subsection{Thermal transport at interfaces}
We finally investigate thermal transport across a coherent Si(001)/diamond(001) interface to test the YB correction under 2D PBCs. The interface was built following Ref.~\cite{rajabpour_sic_2025}, with diamond [110] aligned parallel to Si [110]. The in-plane lattice was constructed by matching three diamond surface periods to two Si periods, and the interface structure was the one with the lowest mean interface excess energy among four termination and registry candidates evaluated with MACE-MH-1 (see SI). The simulation cells have a cross-sectional area of $30.1\times30.1~\text{\AA}^2$ and material lengths ranging from 11.2 to 32.0~nm, with periodic boundary conditions parallel to the interface and vacuum along the transport direction.

To describe the Si/diamond interface, we constructed a Si--C dataset covering crystalline, amorphous, surface, and interfacial environments. Both DPA4C-LR and its short-range counterpart DPA4C were trained on this dataset by distillation from the \texttt{matpes\_r2scan} head of MACE-MH-1 \cite{batatia_macemh_2025}, followed by fine-tuning on r$^2$SCAN DFT data \cite{furness_r2scan_2020}, with the YB correction applied to the long-range electrostatic term for slab configurations. On a test set of 77 configurations sampled from independent interface, surface, and crystal trajectories, DPA4C-LR reaches force RMSEs of 111 meV/{\AA} and energy RMSEs of 5.1 meV/atom against the DFT references (see SI).

We evaluate the interfacial thermal conductance by nonequilibrium molecular dynamics (NEMD) \cite{muller_nemd_1997} for three cells with material lengths of 11, 22, and 32~nm, using the setup of Fig.~\ref{fig:nemd}a. Langevin thermostats \cite{bussi_langevin_2007} at 330 and 270~K establish a steady heat flux, obtained from the energy exchanged with the baths (Fig.~\ref{fig:nemd}b), and the interfacial temperature drop is obtained by extrapolating the temperature profiles on both sides of the interface (Fig.~\ref{fig:nemd}c).

The YB correction is validated against direct 2D Ewald summation \cite{parry_2dewald_1975} on configurations sampled from the NEMD trajectories (Fig.~\ref{fig:nemd}d).
With the YB correction, the 3D Ewald energy agrees with the 2D Ewald reference to within $2\times10^{-9}$~eV at the production vacuum gap of 40.3~{\AA}, whereas the uncorrected 3D Ewald energy deviates by 0.88~eV and retains spurious dipole interactions between periodic images even at a gap of 80~{\AA}.

At material lengths of 11--32~nm, DPA4C-LR gives interfacial thermal conductances of 0.148--0.160~{\GWunit}. 
The DPA4C-LR NEMD agrees in magnitude with a moment-tensor-potential NEMD study of the same interface, which reported about 0.16~{\GWunit} \cite{rajabpour_sic_2025}, and lies slightly above the experimental range of 0.064--0.14~{\GWunit} \cite{cheng_diamondsi_2019,woo_iedm_2023}. For this example, the result demonstrates that DPA4C-LR yields consistent predictions of interfacial thermal conductance with the YB correction implemented effectively.

\section{Discussion}

We show that long-range electrostatics and electrical response can be added to a strictly local MLIP without sacrificing the domain decomposition that makes the MLIP scalable.
We illustrate this combination of accuracy and field coupling with three applications, each grounded in \textit{ab initio} reference data and run at $10^{4}$--$10^{6}$ atoms. These field-driven and thermal-transport simulations are viable because the framework inherits the domain decomposition of the strictly local backbone and the real-space BEC expression of \citet{fan_qnep_2026}.

We find that Eq.~\eqref{eq:BEC} gives a BEC that is both independent of the cell size and distributable across GPUs.
In Ref.~\cite{zhong_les_ir_2025}, LES computes the polarization and BEC via a phase-corrected approach 
\begin{equation}
    \mathbf{Z}^*_i = \Re \left[ \text{diag} \left( \text{e}^{-\text{i}k \mathbf{r}_i} \right) \dfrac{\partial \mathbf{P}(k)}{\partial \mathbf{r}_i } \right], ~\mathbf{P}(k) = \sum_j \frac{ \tilde{q}_j}{\text{i}k}  \text{e}^{\text{i}k\mathbf{r}_j} 
\label{eq:LES_BEC}
\end{equation}
where the exponential is applied componentwise, $\text{diag}(\cdot)$ constructs a diagonal matrix from the components of a vector, $k$ is the magnitude of the smallest reciprocal lattice vector, and $\Re$ denotes the real part.
The LES expression approaches the finite-system definition of the BEC only as $k \rightarrow 0$. In a finite cell, the LES treatment of the BEC therefore induces a finite-size effect. 
This is understandable as the closed form yields
\begin{equation}
    \mathbf{Z}^*_i = \tilde{q}_i \mathbf{I} + \sum_j \frac{\sin (k \mathbf{r}_{ij})}{k} \otimes \frac{\partial \tilde{q}_j}{\partial \mathbf{r}_i}
\end{equation}
where $\mathbf{r}_{ij}=\mathbf{r}_j-\mathbf{r}_i$ is evaluated using the appropriate periodic images, and the sine is applied componentwise. Each displacement component $r$ in the charge-response term is therefore replaced by $\sin(kr)/k$. For $kr\ll1$, the relative attenuation of this geometric factor is approximately $(kr)^2/6$, giving a finite-size correction of order $L^{-2}$ when the local charge response remains unchanged. This does not imply a systematic underestimation of the total BEC, because individual contributions can have different signs. The finite-wavevector error decreases in sufficiently large cells, and replicated supercells were used in Ref.~\cite{zhong_les_ir_2025} to suppress it.

\begin{figure}
    \centering
    \includegraphics[width=0.95\linewidth]{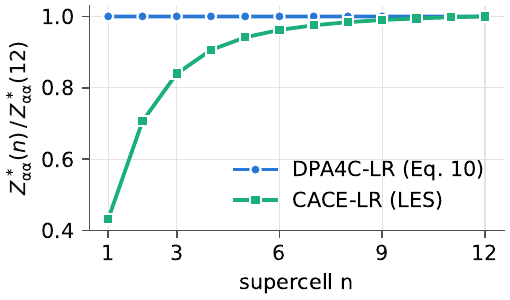}
    \caption{Diagonal BECs ($Z^*_{\alpha\alpha}$) of Nb in a 15-atom \ce{Pb3(Mg_{1}Nb_{2})O9} cell repeated $n\times n\times n$, normalized to $n = 12$, from Eq.~\eqref{eq:BEC} for DPA4C-LR and from Eq.~\eqref{eq:LES_BEC} for CACE-LR with LES.}
    \label{fig:bec_size}
\end{figure}

The real-space expression in Eq.~\eqref{eq:BEC} contains only the relative vectors $\mathbf{r}_{ij}$ and the local derivatives $\partial q^\text{raw}_i / \partial \mathbf{r}_{ij}$, with no reference to a wave vector.
The resulting BEC is therefore independent of the cell size (see Fig.~\ref{fig:bec_size}).
The same expression applies to non-periodic systems, three-dimensional PBC, and slabs, whereas $\mathbf{P}(k)$ is not defined along a non-periodic direction.

The two formulations also differ in how they evaluate the derivatives, which determines whether the long-range augmentation can be parallelized.
In LES, the long-range energy and polarization $\mathbf{P}(k)$ are accumulated over all atoms in the simulation cell, and automatic differentiation yields forces, stresses, and BECs from them. This requires a backward pass through the computational graph of the entire cell.
GPU parallelization of LES-augmented MLIPs is therefore non-trivial, even when the backbone is strictly local.
One viable approach for large systems is graph-based partitioning \cite{kim_data-efficient_2025, han_distmlip_2026}, but it distributes memory across several GPUs and does not necessarily accelerate the simulation.

Here, we differentiate the electrostatic term analytically. The solver returns the direct long-range force and $\partial E^{\text{LR}} / \partial q^\text{raw}_i$, the long-range electrostatic potential at atom $i$ up to a uniform shift.
Each spatial partition then propagates this single scalar per atom through its own charge network, with the same partition, ghost shell, and scatter-sum that the backbone already uses for the short-range forces.
The long-range term adds one global electrostatics solve and one local backward pass per step, and the BEC adds one further local backward pass.
The same factorization applies to any strictly local backbone (e.g., Atomic Cluster Expansion \cite{drautz_atomic_2019}).

The scaling benchmark (Fig.~\ref{fig:water}f--h) quantifies these costs.
For the 1.96-million-atom cell, DPA4C-LR reaches a speed-up of 1.73 on four GPUs, a strong-scaling efficiency of 43\%.
In weak scaling, the parallel efficiency remains at 40\% up to 5.2 million atoms on four GPUs. 
On a single GPU, the long-range term increases the cost of a step by a factor of 2.2 to 2.7 over the short-range DPA4C backbone.
Computing the BEC together with the forces adds less than 20\% to a force evaluation (Fig.~\ref{fig:water}h vs.\ g).

We acknowledge two limitations of the current framework. First, the receptive field of the strictly local descriptor is limited to a single cutoff radius, which may restrict its ability to capture medium-range interactions compared with message-passing architectures such as NequIP \cite{batzner_nequip_2022} and MACE \cite{batatia_mace_2022}. Nevertheless, previous benchmarks show that strictly local models such as Allegro \cite{musaelian_learning_2023} can achieve competitive accuracy \cite{kozinsky_scaling_2023}, especially when augmented with long-range electrostatics \cite{kim_les_2025}. Consistent with these findings, DPA4C-LR reaches a force RMSE of 20.2~meV/{\AA} and a BEC RMSE of 0.05~$e$ for bulk water. 
These results suggest that explicitly incorporating long-range electrostatics can improve the predictive and inference capabilities of strictly local MLIPs without additional message passing.
Second, the locally predicted latent charges cannot fully capture nonlocal charge redistribution, as occurs in metallic systems or at metal--electrolyte interfaces. Describing such effects generally requires global charge equilibration, which introduces additional computational and communication costs \cite{ko_fast_2025, gao_foundation_2025, wang_ion-modulated_2026}.
While we include only monopoles in this work, latent dipoles and quadrupoles can also be read out from local descriptors, so the same factorization should carry over to multipolar electrostatics and polarizability-dependent observables such as Raman spectra \cite{kim_polarizable_2026}.
Our approach is therefore best suited to system-specific, device-scale applications, where distillation and fine-tuning provide an accurate and scalable potential for the target chemistry.
For foundation models, all-to-all attention can learn long-range interactions directly from data without explicit physical terms \cite{qu_recipe_2026}, offering an alternative approach that leverages the scalability of attention-based architectures \cite{vaswani_attention_2017}.

Looking ahead, with the spatial parallelization of BEC inference, we also envision further acceleration through hardware-aware GPU execution and non-von Neumann architectures, such as field-programmable gate arrays (FPGAs), for massive-scale simulations under external fields (e.g., direct device simulations with millions to billions of atoms). A strictly local model reduces to spatially localized per-atom kernels that map naturally onto parallel hardware, with global coupling confined to charge constraints, collective reductions, and a fast Fourier transform-based long-range electrostatics solve. 
Recent advances in GPU inference, exemplified by Symmetrix-XL, show that architecture-preserving execution can extend the accessible scale of pretrained equivariant potentials \cite{fu_symmetrix_xl_2026}. A hardware co-designed deep neural network potential deployed on a single FPGA has been reported to reach about $10^{-7}$~s per step per atom for systems of up to 20 million atoms \cite{yu_nvnmd-v2_2025}. 
Combined with long-range electrostatics and BEC inference, such advances could extend atomistic simulations with \textit{ab initio} accuracy under an applied electric field to device layer scales.

In summary, we develop a scalable approach that incorporates long-range electrostatics and electrical response into strictly local equivariant MLIPs while preserving spatial decomposability. Distributed evaluation of forces and Born effective charges extends electrical-response modeling to multi-million-atom systems. This advance provides a practical foundation for investigating field-driven and interfacial dynamics, bringing atomistic materials modeling closer to device-relevant scales.

\section*{Acknowledgments} This work was supported in part by the AI2050 program at Schmidt Sciences (Grant G-25-69776), NUS-AISI Joint Research Initiative Fund, and computational resources from the NUS High Performance Computing facility (NUSREC-HPC-00001). The authors thank Yubai Shi for providing the PMN training dataset.

\section*{Data availability}
The supporting data and code are available at \url{https://github.com/AM3GroupHub/deepmd-lr}. 

\section*{Author contributions}
P.Z. conceived and designed the research. Y.C. developed the model and codebase. X.Y. contributed to workflow automation.
H.S., Z.L., X.W. and Z.X. performed simulations and research with help from Y.C. and J.C. 
Y.C. and P.Z. wrote the manuscript with input from others. 
All authors contributed to discussions.

\section*{Competing interests}
The authors declare no competing interests.

\bibliography{references}
\end{document}